\documentclass[aps,prd,floats,superscriptaddress,altaffilletter]{revtex4}
\usepackage{amsmath,amssymb,graphicx}

\newcommand{\sgn}{\operatorname{{\mathrm sgn}}}

\begin{document}

\newcommand{\be}{\begin{equation}}
\newcommand{\ee}{\end{equation}}
\newcommand{\bea}{\begin{eqnarray}}
\newcommand{\eea}{\end{eqnarray}}
\def \wc {w_{c,\text{eff}}}
\def \wv {w_{\varphi,\text{eff}}}

\title{Quintessence with quadratic coupling to dark matter}

\author{Christian G. B\"ohmer}
\email{c.boehmer@ucl.ac.uk}
\affiliation{Department of Mathematics and Institute of Origins, 
University College London, London WC1E 6BT, UK}

\author{Gabriela Caldera-Cabral}
\email{gaby.calderacabral@port.ac.uk}
\affiliation{Institute of Cosmology \& Gravitation, University of
Portsmouth, Portsmouth PO1 3FX, UK}

\author{Nyein Chan}
\email{nyein.chan@ucl.ac.uk}
\affiliation{Department of Mathematics and Institute of Origins, 
University College London, London WC1E 6BT, UK}

\author{Ruth Lazkoz}
\email{ruth.lazkoz@ehu.es}
\affiliation{Fisika Teorikoa, Euskal Herriko Unibertsitatea, 48080
Bilbao, Spain}

\author{Roy Maartens}
\email{roy.maartens@port.ac.uk}
\affiliation{Institute of Cosmology \& Gravitation, University of
Portsmouth, Portsmouth PO1 3FX, UK}

\date{\today}

\begin{abstract}

We introduce a new form of coupling between dark energy and dark
matter that is quadratic in their energy densities. Then we
investigate the background dynamics when dark energy is in the
form of exponential quintessence. The three types of quadratic coupling all admit late-time accelerating critical points, but these are not scaling solutions. We also show that two types of coupling allow for a suitable matter era at early times and acceleration at late times, while the third type of coupling does not admit a suitable matter era.

\end{abstract}

\maketitle

\section{Introduction}

Cosmological observations strongly suggest that the expansion rate
of the universe is accelerating and that matter in the universe is
dominated by non-baryonic cold dark matter (see
e.g.~\cite{Dunkley:2008ie}). However, what exactly causes this
acceleration is not well understood, and one of the main
challenges of modern cosmology is to understand the nature of this
mysterious dark energy. The existence of some form of dark matter
is long known, as implied by the flattened galactic rotation
curves observed by Zwicky as early as 1933. Several experiments
have been carried out (see, for
example~\cite{Collaboration:2009nf}) in search of candidate dark
matter particles. The fact that dark matter only interacts weakly
with standard matter means that it is difficult to detect such
particles directly. Neither dark energy nor dark matter have
been detected directly. Only the total dark sector 
energy-momentum tensor is known
from its combined gravitational effect. In order to separate the
two components, we have to assume a model for them. It is possible
that these components interact with each other, while not being coupled
to standard model particles. Such a possibility can lead to new
approaches to the coincidence problem (``how do dark matter and
dark energy attain the same order-of-magnitude value at the right
time to allow for the observed large-scale structure?"). It can
also produce interesting new features in large-scale structure,
such as a large-scale gravitational bias~\cite{Amendola:2001rc}
and a violation of the weak equivalence principle by dark matter
on cosmological scales~\cite{Koyama:2009gd}.

In this paper we study a class of cosmological models with
interactions in the dark sector. Various models of the coupling
between dark energy and dark matter have been proposed and
investigated (see, e.g.~\cite{Wetterich:1994bg,Amendola:1999qq,Billyard:2000bh,Zimdahl:2001ar, Farrar:2003uw,Chimento:2003iea,Olivares:2005tb, Sadjadi:2006qp,Guo:2007zk,Kim:2007dp,Boehmer:2008av,He:2008tn, Chen:2008ca,Quartin:2008px, Pereira:2008at,Quercellini:2008vh,Valiviita:2008iv}). We consider
only the background dynamics -- for cosmological perturbations of
coupled dark energy models, see e.g.~\cite{Amendola:2002bs,Koivisto:2005nr,Olivares:2006jr,Mainini:2007ft, Bean:2007ny,Valiviita:2008iv,Vergani:2008jv,Pettorino:2008ez, Schaefer:2008qs,Schaefer:2008ku,LaVacca:2008kq,He:2008si,Bean:2008ac, Corasaniti:2008kx,Chongchitnan:2008ry,Jackson:2009mz,Gavela:2009cy, LaVacca:2009yp,He:2009mz,CalderaCabral:2009ja,He:2009pd, Koyama:2009gd,Valiviita:2009nu,Majerotto:2009np}.

There is no fundamental theory that selects a specific coupling in
the dark sector, and therefore any coupling model will necessarily
be phenomenological, although some models will have more physical
justification than others. Here we analyse the background dynamics
for a new model of coupling. This model improves the one
previously introduced in~\cite{Boehmer:2008av,Valiviita:2008iv},
which was motivated by simple models of inflaton decay during
reheating and of curvaton decay to radiation.

The background description of a coupled model with quintessence
dark energy density $\rho_{\varphi}$ and dark matter density
$\rho_c$ is given by the energy balance equations
\begin{align}
  \dot \rho_c  &= - 3H\rho_c+Q_c\,,
  \label{cc}\\
  \dot \rho_{\varphi} &= - 3H(1+w_{\varphi})\rho_{\varphi}+
Q_\varphi\,, \qquad Q_\varphi=-Q_c := Q \,.
  \label{kg1}
\end{align}
Here $Q_A$ is the rate of energy transfer to species $A$. It
follows that
 \begin{equation} \label{sq}
   Q~\left\{
   \begin{array}{l}
     >0\\ <0
   \end{array} \right.\quad
   \Rightarrow \quad \mbox{energy transfer is}~\left\{
   \begin{array}{l}
\mbox{dark matter $\to$ dark energy}\\ \mbox{dark energy $\to$
dark matter}
   \end{array} \right.
 \end{equation}
The dark energy equation of state parameter is
\begin{equation}
w_\varphi:=\frac{p_\varphi}{\rho_\varphi}=\frac{\frac{1}{2}
\dot\varphi^2-V(\varphi)}
   {\frac{1}{2}\dot\varphi^2+V(\varphi)}\,.\label{wp}
\end{equation}
The modified Klein-Gordon equation follows from Eq.~(\ref{kg1}) as
\begin{equation}
\ddot \varphi +3H \dot \varphi + \frac{dV}{d\varphi}=
\frac{Q}{\dot\varphi}\,.
  \label{kg}
\end{equation}
For quintessence with an exponential potential,
\begin{equation}
V(\varphi) = V_0 \exp \left(-\kappa \lambda \varphi\right)\,,\qquad
  \kappa^2:=8\pi G\,,
\end{equation}
where $\lambda$ is a dimensionless parameter and $V_0>0$. We neglect the radiation and therefore the evolution equations are
\begin{align}
\dot \rho_b &= - 3H\rho_b \,,\\
\dot H &= -\frac{\kappa^2}{2}\left[\rho_c+\rho_b+\dot\varphi^2
\right]\,,
  \label{ray}
\end{align}
where baryons $\rho_b$ are not coupled to the dark sector. The
Friedman constraint is
\begin{align}
  \label{fried}
  H^2= \frac{\kappa^2}{3}\left(\rho_c+\rho_b+\rho_{\varphi}\right).
\end{align}

We define effective equation of state parameters for the dark
components which describe the equivalent uncoupled model in the
background: $\dot \rho_c +3H(1+ \wc)\rho_c=0 $, $\dot \rho_\varphi
+3H(1+ \wv)\rho_\varphi=0 $. By Eqs.~(\ref{cc}) and~(\ref{kg1}),
\begin{equation}
  \label{weff}
\wc = \frac{Q}{3H\rho_c}\,,\qquad \wv= w_\varphi -
\frac{Q}{3H\rho_\varphi}\,.
\end{equation}

\section{New model of dark sector coupling}

In~\cite{Boehmer:2008av,Valiviita:2008iv} a model of the form
\begin{equation}
  Q=\Gamma \rho_c,
  \label{old}
\end{equation}
was introduced, with $\Gamma$ constant. The motivation for this
form of interaction is that, for $\Gamma>0$, the same $Q$ is used
for simple models of: (1)~the decay of an inflaton field to
radiation during reheating~\cite{Turner:1983he}, (2)~the decay of
dark matter into radiation~\cite{Cen:2000xv}, (3)~the decay of a
curvaton field into radiation~\cite{Malik:2002jb}, (4)~the decay
of super-heavy dark matter particles into a scalar
field~\cite{Ziaeepour:2003qs}.

We consider this coupling to be better motivated than alternatives
of the form $Q=\alpha H \rho_c$ -- which are designed for
mathematical simplicity, since they lead to the same number of
dimensions of the phase space (two) as the uncoupled case. The
coupling in Eq.~(\ref{old}) is not designed for mathematical
simplicity, but is chosen as a physically simple form of decay
law. It leads to a three-dimensional phase space. This new phase
can be compactified~\cite{Boehmer:2008av}, as in the
two-dimensional case, but great care is required in analysing the
stability properties of the resulting dynamical system. The
stability matrix contains singular eigenvalues as one approaches
the critical points. In~\cite{Boehmer:2008av} we developed the
required machinery to overcome these problems and were able to
present a complete phase space analysis. Our techniques are
readily applicable to more general couplings.

Simple decay laws of the form in Eq.~(\ref{old}) fail to reflect
the feature that interactions are typically determined by {\em
both} energy densities. We therefore consider the natural first
extension Eq.~(\ref{old}) to a quadratic form
\begin{equation}
Q=\mathcal{A} \rho_\varphi^2 + \mathcal{B} \rho_c^2 + \mathcal{C}
\rho_c \rho_\varphi\,,\label{new}
\end{equation}
where $\mathcal{A}$, $\mathcal{B}$ and $\mathcal{C}$ are coupling
constants. We define dimensionless coupling constants as
\begin{equation}
\alpha= \mathcal{A}{H_0} \,, ~~ \beta= \mathcal{B}{H_0}\,,~~ \gamma=\mathcal{C}{H_0}
\,.
\end{equation}

The Friedman constraint~(\ref{fried}) in dimensionless form
becomes
\begin{equation}
  \label{constr}
\Omega_c+\Omega_\varphi = 1\,,\qquad \Omega :=
\frac{\kappa^2\rho}{3H^2}\,,
\end{equation}
where we neglect the baryons, and the total equation of state 
parameter is given by
\begin{align}
  \label{tot}
&\dot\rho_{\text{tot}} + 3 H (1+w_{\text{tot}})
\rho_{\text{tot}}=0\,,\\
&w_{\text{tot}}:= \frac{p_{\text{tot}}}{\rho_{\text{tot}}} =
\frac{p_\varphi}{\rho_\varphi+\rho_c} = w_\varphi\Omega_\varphi\,.
\end{align}
The condition for acceleration is $w_{\text{tot}}<-1/3$. A phantom field with $w_{\varphi} <-1$ violates the dominant energy condition, $\rho\geq |p|$. We therefore assume that $w_{\varphi}>-1$, thereby excluding phantom models with negative kinetic energy.

We introduce the dimensionless variables $x,y$, as in the
uncoupled case~\cite{Copeland:1997et}, where
\begin{align}
  x^2 := \frac{\kappa^2 \dot {\varphi}^2}{{6}H^2}\,,\qquad
  y^2 := \frac{\kappa^2 {V}}{{3}H^2}\,.
  \label{def1}
\end{align}
Then  $y \geq 0$ because of the positivity of the potential
energy, and Eq.~(\ref{constr}) implies that
\begin{equation}
  0\leq\Omega_\varphi=x^2+y^2 \leq 1\,.
\end{equation}
In the new variables, the equation of state parameters are
\begin{equation}
w_\varphi= \frac{x^2-y^2}{x^2+y^2}\,,\qquad w_{\text{tot}} =
x^2-y^2\,.
\end{equation}
The Hubble evolution equation may be written as
\begin{equation}
\frac{ \dot H}{H^2}=- \frac{3}{2}(1+x^2-y^2)\,.
  \label{reH}
\end{equation}
As already indicated, it turns out that the resulting evolution
equations do not allow a two-dimensional representation of this
model, since we cannot eliminate $H$ from the energy balance
equations~(\ref{cc}) and (\ref{kg1}), using only the variables
$x(N),y(N)$, where we use $N=\log(a)$ as the independent variable. 
Equation~(\ref{reH}) must therefore be incorporated
into the dynamical system. We do this via a new variable $z$,
chosen so as to maintain compactness of the phase space
\begin{equation}
  z = \frac{H_0}{H+H_0}\,.
  \label{change2}
\end{equation}
Thus $0\leq z \leq 1$, and the compactified phase space now
corresponds to a half-cylinder of unit height and
radius~\cite{Boehmer:2008av}. The top of this half-cylinder is
defined by $z=1$ and it turns out that the equations become
singular as $z \to 1$. Therefore, care is required in order to
analyse the resulting dynamical system.

It is evident that the quadratic and higher-order couplings can be
treated in a similar fashion, for instance one could consider a
coupling of the form $Q = \mu \rho_\varphi^3\rho_c$. The most
general model is of the form
\begin{align}
  Q = \sum\limits_{m,n} q_{mn}\rho_c^m \rho_\varphi^n\,,
\end{align}
where $m,n$ are non-negative integers. The matrix $q_{mn}$ is
arbitrary except for the condition $q_{00}=0$. Note that $q_{mn}$
has no a priori symmetry properties and is not necessarily a
square matrix. The linear model in Eq.~(\ref{old}) was extended to
the most general linear model, $Q=-(\Gamma_c\rho_c +
\Gamma_\phi\rho_\phi)$ in~\cite{CalderaCabral:2008bx}. The linear
and quadratic models lead to
\begin{align}
  (q)_{mn} = \begin{pmatrix}  0 & \Gamma_\phi \\
  \Gamma_c &  0  \end{pmatrix},~~~
  (q)_{mn} = \begin{pmatrix} 0 & 0 & \mathcal{A} \\
  0 & \mathcal{C} & 0 \\ \mathcal{B} & 0 & 0 \end{pmatrix}.
\end{align}

\section{Dynamical analysis}

In this section we analyse the three particular cases when two of
the interaction terms are equal to zero. Two of these models allow
for a standard matter era, but the model with
$\mathcal{A}=\mathcal{C}=0$ does not allow it. Then we combine the
models $\mathcal{A}$ and $\mathcal{C}$ and analyse the composite
model.

\subsection{Model $\mathcal{A}$: coupling $Q = \frac{\alpha}{H_0} \rho_\varphi^2$}

The system of autonomous differential equations is
\begin{align}
  \label{ax'}
  x'&=-3x+\lambda \frac{\sqrt{6}}{2}\,y^2+\frac{3}{2}x(1+x^2-y^2)
  +\alpha\,\frac{3(1-z)(x^2+y^2)^2}{2xz}\,,\\
  \label{ay'}
  y'&=-\lambda \frac{\sqrt{6}}{2}\,xy + \frac{3}{2}y(1+x^2-y^2)\,,\\
  \label{az'}
  z'&= \frac{3}{2}z(1-z)(1+x^2-y^2)\,.
\end{align}
The critical points, defined by $x'=0,y'=0$ and $z'=0$, and the
eigenvalues of the stability matrix are given in
Table~\ref{model1eigen}. In Table~\ref{model1stab} we characterize
the critical points and give the effective equation of state for
the late-time attractor.

\begin{table}[!htb]
\begin{tabular}[t]{|l|c|c|c|c|}
\hline
Point & $x_*$ & $y_*$ & $z_*$ & Eigenvalues \\[1ex]
\hline
\hline &&& \\[-3ex]
A & $0$ & $0$ & $0$ & $-\frac{3}{2}\,, \frac{3}{2}\,, \frac{3}{2}$ \\[1ex]
\hline
\hline &&& \\[-3ex]
D & $0$ & $0$ & $1$ & $-\frac{3}{2}\,, -\frac{3}{2}\,, \frac{3}{2}$ \\[1ex]
\hline &&& \\[-3ex]
E$_{\pm}$ & $\pm 1$ & $0$ & $1$ & $-3\,, 3\,, 3\mp\sqrt{\frac{3}{2}}\lambda$ \\[1ex]
\hline
\hline &&& \\[-3ex]
F & $\sqrt{\frac{3}{2}}\frac{1}{\lambda}$
&$\sqrt{\frac{3}{2}}\frac{1}{\lambda}$  & 1 &
$-\frac{3}{2}\,, -\frac{3}{4\lambda}(\lambda\pm\sqrt{24-7\lambda^2})$\\[1ex]
\hline &&& \\[-3ex]
G & $\frac{\lambda}{\sqrt{6}}$ & $\sqrt{1-\frac{\lambda^2}{6}}$ &
1 &
$-\frac{\lambda^2}{2}\,, -3+\frac{\lambda^2}{2}\,, -3+\lambda^2$ \\[1ex]
\hline
\end{tabular}
\caption{Critical points and associated eigenvalues for coupling
model $\mathcal{A}$.} \label{model1eigen}
\end{table}

\begin{table}[!htb]
\begin{tabular}[t]{|l|c|c|c|c|c|c|}
\hline
Point & Stable?& $\Omega_{\phi}$ & $w_T$& Acceleration?& Existence \\
\hline \hline &&&&& \\[-3ex]
A & Saddle node & 0 & 0 &No & $\forall \, \lambda,\,\alpha $ \\[1ex]
\hline
\hline &&&&& \\[-3ex]
D & Saddle node& 0 & 0 &No& $\forall \, \lambda,\,\alpha $\\[1ex]
\hline &&&&& \\[-3ex]
E$_{\pm}$ & Saddle node & 1 & 1 & No &$\forall \, \lambda,\,\alpha $\\[1ex]
\hline &&&&& \\[-3ex]
F & Stable focus for $\lambda^2 >\frac{24}{7}$
&$\frac{3}{\lambda^2}$ & 0&No &
$\lambda ^2 > 3$\\[1ex]
   &Stable node for $3<\lambda^2 <\frac{24}{7}$&&&& \\[1ex]
\hline &&&&&\\[-3ex]
G & Saddle node for $\lambda^2>3$ & 1& $\frac{\lambda^2}{3}-1$ &
$\lambda^2<2$ &
$\lambda^2 <6$\\[1ex]
 &Stable node for $\lambda^2<3$&&&& \\[1ex]
\hline
\end{tabular}
\caption{The properties of the critical points for model
$\mathcal{A}$.} \label{model1stab}
\end{table}

\begin{figure}[!htb]
\centering
\includegraphics[width=0.50\textwidth]{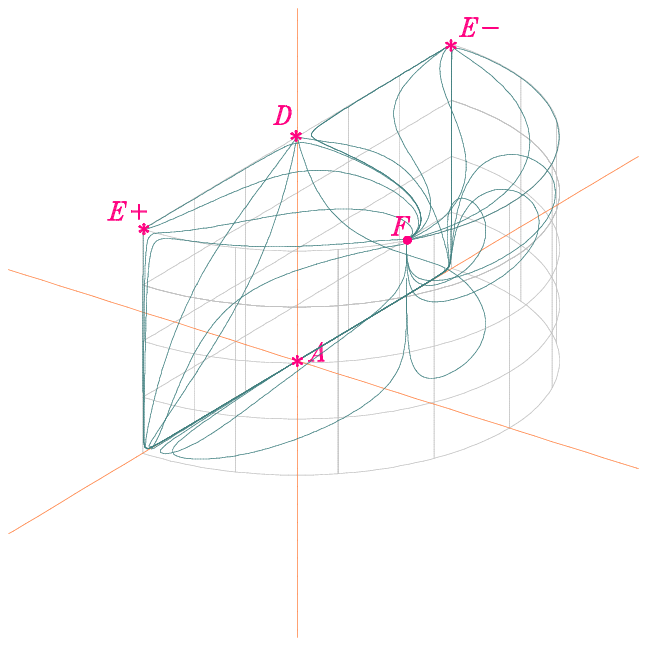}\hfill
\includegraphics[width=0.50\textwidth]{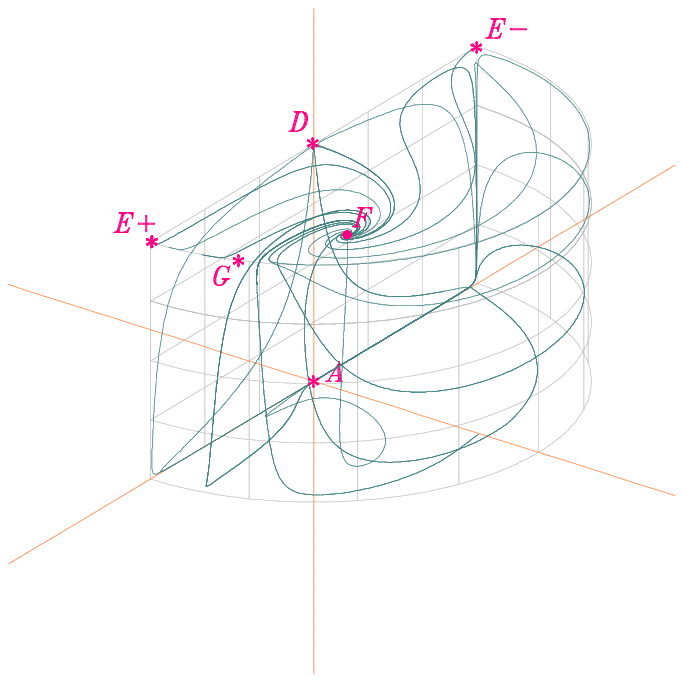}
\caption{Phase-space trajectories for model $\mathcal{A}$. The
left plot shows the stable node G, with $\lambda=1.2$ and
$\alpha=10^{-3}$. The right plot shows the stable focus F with
$\lambda=2.3$ and $\alpha=10^{-3}$ } \label{model1}
\end{figure}

This model depicts  an evolution of the universe in good agreement with the observations for certain values
of the parameters $\lambda, \alpha$. Saddle
point A corresponds to the standard matter dominated universe with
$a(t) \propto t^{2/3}$; its instability allows for the existence of  trajectories escaping from it and  ending at 
an attractor, which exists for adequate values of the parameters.  However, the attractor (or late time stage of the universe) will only represent an accelerated scenario for a flat enough potential,
specifically when $\lambda^2<2$. 
In this case the attractor is completely dark energy dominated, point G.
If the potential is not flat enough, the attractor, point F, is a scaling solution, in which the fraction
of dark energy is the dominant one.

One way to understand this behaviour is by looking at the relative
strength of the coupling, $f$, in the CDM balance equation,
\begin{equation}
  f:=\frac{|Q|}{H \rho_c}\,.
  \label{strength}
\end{equation}
In the matter era, $  H^2={\rho_c}/{3},$ we have
\begin{equation}
  f\sim \rho_{\phi}^2/\rho_c^{3/2},
\end{equation}
which is decreasing into the past. Thus the coupling is weaker in
the past and this allows a near-standard matter era. It is
important to note that the dynamical behaviour of the standard matter point A and the
accelerated critical solution G does not depend on the sign of the
coupling parameter $\alpha$.

\subsection{Model $\mathcal{B}$: coupling $Q=\frac{\beta}{H_0}\rho_{c}^2$}

The autonomous system is
\begin{align}
  \label{bx'}
  x'&=-3x+\lambda \frac{\sqrt{6}}{2}\,y^2+\frac{3}{2}x(1+x^2-y^2)
  +\beta\,\frac{3(1-z)(1-x^2-y^2)^2}{2xz}\,,\\
  \label{by'}
  y'&=-\lambda \frac{\sqrt{6}}{2}\,xy + \frac{3}{2}y(1+x^2-y^2)\,,\\
  \label{bz'}
  z'&= \frac{3}{2}z(1-z)(1+x^2-y^2)\,.
\end{align}
The critical points and their stability properties are summarized
in Tables~\ref{modelIIeigen} and~\ref{modelIIstab} respectively.

\begin{table}[!htb]
\begin{tabular}[t]{|l|c|c|c|c|c|c|}
\hline
Point & $x_*$ & $y_*$ & $z_*$ & Eigenvalues & $w_{tot}$\\
\hline \hline &&&& \\[-3ex]
B$_{\pm}$ & $\pm 1$ & $0$ & $0$ & $3\,,3\,,3\mp\sqrt{\frac{3}{2}}
\lambda$ & 1\\[1ex]
\hline
\hline &&&& \\[-3ex]
C & $\frac{\lambda}{\sqrt{6}}$ & $\sqrt{1-\frac{\lambda^2}{6}} $ & 0 &
$\frac{\lambda^2}{2}\,,\frac{\lambda^2}{2}-3\,,\lambda^2-3$ &
$\frac{\lambda^{2}}{3}-1$\\[1ex]
\hline
\hline &&&& \\[-3ex]
D & $0$ & $0$ & $1$ & $-\frac{3}{2}\,,-\frac{3}{2}\,,\frac{3}{2}$ & 0\\[1ex]
\hline &&&& \\[-3ex]
E$_{\pm}$ & $\pm 1$ & $0$ & $1$ & $-3\,,3\,,3\mp\sqrt{\frac{3}{2}}
\lambda$ & 1\\[1ex]
\hline
\hline &&&& \\[-3ex]
F & $\sqrt{\frac{3}{2}}\frac{1}{\lambda}$
&$\sqrt{\frac{3}{2}}\frac{1}{\lambda}$  & 1 &
$-\frac{3}{2}\,, -\frac{3}{4\lambda}(\lambda\pm\sqrt{24-7\lambda^2})$ & 0\\[1ex]
\hline &&&& \\[-3ex]
G & $\frac{\lambda}{\sqrt{6}}$ & $\sqrt{1-\frac{\lambda^2}{6}}$ & 1 &
$-\frac{\lambda^2}{2}\,,-3+\frac{\lambda^2}{2}\,,-3+\lambda^2$ &
$\frac{\lambda^{2}}{3}-1$\\[1ex]
\hline
\end{tabular}
\caption{Critical points and associated eigenvalues for coupling
model $\mathcal{B}$.} \label{modelIIeigen}
\end{table}

\begin{table}[!htb]
\begin{tabular}[t]{|l|c|c|c|c|c|c|}
\hline
Point & Stable?& $\Omega_{\phi}$ & $w_T$& Acceleration?& Existence \\
\hline \hline &&&&& \\[-3ex]
B$_{+}$ & Saddle node for $\lambda >\sqrt{6}$ & 1 & 1 &No & $\forall \,
\lambda,\,\beta $ \\[1ex]
 &  Unstable node for $\lambda <\sqrt{6}$  &&&&\\
\hline
\hline &&&&& \\[-3ex]
B$_{-}$ & Unstable node for $\lambda >-\sqrt{6}$ & 1 & 1 &No & $\forall \,
\lambda,\,\beta $ \\[1ex]
 &  Saddle node for $\lambda <-\sqrt{6}$  &&&&\\
\hline
\hline &&&&& \\[-3ex]
C & Saddle node& 1 & $\frac{\lambda^2}{3}-1$ &$\lambda^2<2$& $\lambda^2<6 $\\[1ex]
\hline &&&&& \\[-3ex]

D & Saddle node& 0 & 0 &No& $\forall \, \lambda,\,\beta $\\[1ex]
\hline &&&&& \\[-3ex]
E$_{\pm}$ & Saddle node & 1 & 1 & No &$\forall \, \lambda,\,\beta $\\[1ex]
\hline &&&&& \\[-3ex]
F & Stable focus for $\lambda^2 >\frac{24}{7}$
&$\frac{3}{\lambda^2}$ & 0&No &
$\lambda ^2 > 3$\\[1ex]
   &Stable node for $3<\lambda^2 <\frac{24}{7}$&&&& \\[1ex]
\hline &&&&&\\[-3ex]
G & Saddle node for $\lambda^2>3$ & 1& $\frac{\lambda^2}{3}-1$ &
$\lambda^2<2$ &
$\lambda^2 <6$\\[1ex]
 &Stable node for $\lambda^2<3$&&&& \\[1ex]
\hline
\end{tabular}
\caption{The properties of the critical points for model
$\mathcal{B}$.} \label{modelIIstab}
\end{table}

\begin{figure}[!htb]
\centering
\includegraphics[width=0.50\textwidth]{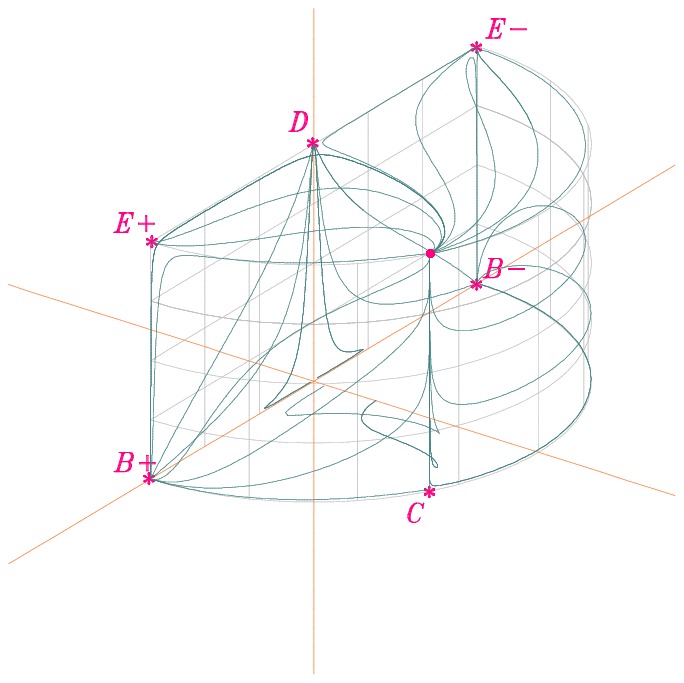}\hfill
\includegraphics[width=0.50\textwidth]{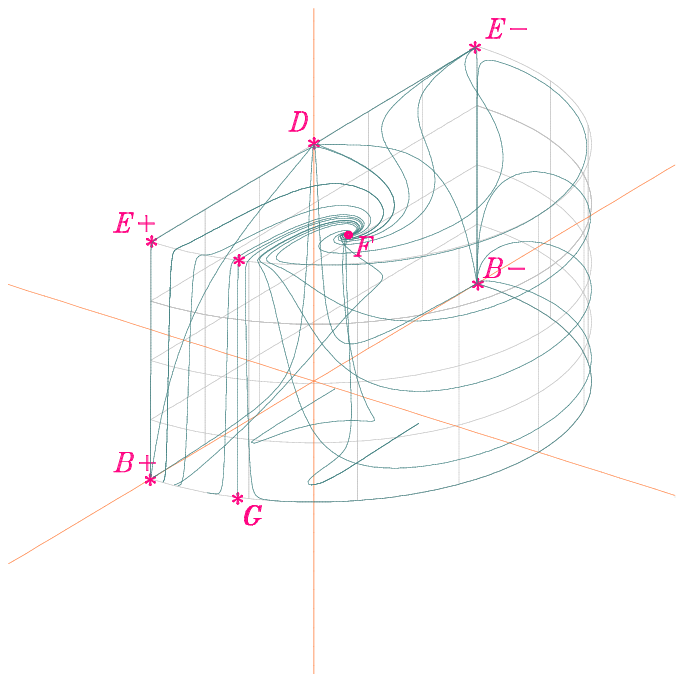}
\caption{Phase-space trajectories for model $\mathcal{B}$. The
left plot shows the stable node G, with $\lambda=1.2$ and
$\beta=10^{-3}$. The right plot shows the stable focus F with
$\lambda=2.3$ and $\beta=10^{-3}$. } \label{model2}
\end{figure}

This model does not have a unstable standard matter solution, so it cannot 
depict an evolution from an early dust-like scenario. However, there is a solution in which the dark energy can
mimic such behaviour for $\lambda^2=3$.   This is saddle point C, but the $\lambda$ value
allowing for that point  to act like matter prevents the existence of  an accelerated
solution at late times. Regarding the late-time attractor we meet again the same situation as before, a scaling but dark energy dominated non accelerated solutions for excessively shallow potentials, and an accelerated completely dark energy dominated scenario
in the opposite case. The relative strength of the coupling
Eq.~(\ref{strength}) for this model is
\begin{equation}
f\sim H \,,
\end{equation}
and is increasing to the past. Since the coupling gets stronger at
early times we cannot get a standard matter era. The direction of
the energy exchange does not affect this conclusion.

\subsection{Model $\mathcal{C}$: coupling $ Q=\frac{\gamma}{H_0} \rho_c \rho_\varphi$}

In this case,
\begin{align}
  \label{cx'}
  x'&=-3x+\lambda \frac{\sqrt{6}}{2}\,y^2+\frac{3}{2}x(1+x^2-y^2)
  +\gamma\,\frac{3(1-z)(1-x^2-y^2)(x^2+y^2)}{2xz}\,,\\
  \label{cy'}
  y'&=-\lambda \frac{\sqrt{6}}{2}\,xy + \frac{3}{2}y(1+x^2-y^2)\,,\\
  \label{cz'}
  z'&= \frac{3}{2}z(1-z)(1+x^2-y^2)\,,
\end{align}
and we summarize the critical points and stability in
Tables~\ref{modelIIIeigen} and~\ref{modelIIIstab}.

\begin{table}[!htb]
\begin{tabular}[t]{|l|c|c|c|c|c|}
\hline
Point & $x_*$ & $y_*$ & $z_*$ & Eigenvalues & $w_{tot}$\\
\hline \hline &&&& \\[-3ex]
A & $0$ & $0$ & $0$ & $\frac{3}{2}\,, \frac{3}{2}\,, \sgn(\gamma)\infty$ &0\\[1ex]
\hline &&&& \\[-3ex]
B$_{\pm}$ & $\pm 1$ & 0 & 0 & $3\,, 3\mp\sqrt{\frac{3}{2}}\lambda\,,
-\sgn(\gamma)\infty$ & 1 \\[1ex]
\hline &&&&\\[-3ex]
C & $\frac{\lambda}{\sqrt{6}}$ & $\sqrt{1-\frac{\lambda^2}{6}}$ & 0 &
$\frac{\lambda^2}{2}\,, \frac{\lambda^2}{2}-3\,,-\sgn(\gamma)\infty$ & 1\\[1ex]
\hline
\hline &&&& \\[-3ex]
D & 0 & 0 & 1 & $-\frac{3}{2}\,, -\frac{3}{2}\,,\frac{3}{2}$ & 0\\[1ex]
\hline &&&& \\[-3ex]
E$_{\pm}$ & $\pm 1$ & 0 & 1 & $-3\,, 3\,,3\mp\sqrt{\frac{3}{2}}\lambda$ & 1\\[1ex]
\hline
\hline &&&& \\[-3ex]
F & $\sqrt{\frac{3}{2}}\frac{1}{\lambda}$ & $\sqrt{\frac{3}{2}} \frac{1}{\lambda} $
& 1 & $-\frac{3}{2}\,, -\frac{3}{4\lambda}(\lambda\pm\sqrt{24-7\lambda^2}) $
 & 0\\[1ex]
\hline &&&& \\[-3ex]
G & $\frac{\lambda}{\sqrt6}$ & $\sqrt{1-\frac{\lambda^2}{6}}$ & 1 &
$-\frac{\lambda^2}{2}\,,
 -3+\frac{\lambda^2}{2}\,,-3+\lambda^2$ & $\frac{\lambda^{2}}{2}-1$\\[1ex]
\hline \hline
\end{tabular}
\caption{Critical points and associated eigenvalues for coupling
model $\mathcal{C}$.} \label{modelIIIeigen}
\end{table}

\begin{table}[!htb]
\begin{tabular}[t]{|l|c|c|c|c|c|c|}
\hline
Point & Stable?& $\Omega_{\phi}$ & $w_T$& Acceleration?& Existence \\
\hline \hline &&&&& \\[-3ex]
A & Unstable node for $\gamma>0$ & 0 & 0 &No & $\forall \, \lambda,\,
\gamma $ \\[1ex]
 &  Saddle  node for $\gamma<0$  &&&&\\
\hline
\hline &&&&& \\[-3ex]
B$_{+}$ & Unstable node for$\lambda <\sqrt{6}$ and $\gamma<0$ & 1 & 1 &No
& $\forall \, \lambda,\,\gamma $ \\[1ex]
 & Saddle otherwise  &&&&\\
\hline
\hline &&&&& \\[-3ex]
B$_{-}$ & Unstable node for $\lambda >-\sqrt{6}$ and $\gamma <0$ & 1 & 1 &No &
$\forall \, \lambda,\,\gamma $ \\[1ex]
 &  Saddle otherwise  &&&&\\
\hline
\hline &&&&& \\[-3ex]
C & Saddle node & 1 & $\frac{\lambda^2}{3}-1$ & $\lambda^2<2$& $\lambda^2<6 $\\[1ex]
\hline &&&&& \\[-3ex]
D & Saddle node& 0 & 0 &No& $\forall \, \lambda,\,\beta $\\[1ex]
\hline &&&&& \\[-3ex]
E$_{\pm}$ & Saddle node & 1 & 1 & No &$\forall \, \lambda,\,\beta $\\[1ex]
\hline &&&&& \\[-3ex]
F & Stable focus for $\lambda^2 >\frac{24}{7}$
&$\frac{3}{\lambda^2}$ & 0&No &
$\lambda ^2 > 3$\\[1ex]
   &Stable node for $3<\lambda^2 <\frac{24}{7}$&&&& \\[1ex]
\hline &&&&&\\[-3ex]
G & Saddle node for $\lambda^2>3$ & 1& $\frac{\lambda^2}{3}-1$ &
$\lambda^2<2$ &
$\lambda^2 <6$\\[1ex]
 &Stable node for $\lambda^2<3$&&&& \\[1ex]
\hline
\end{tabular}
\caption{The properties of the critical points for model
$\mathcal{C}$.} \label{modelIIIstab}
\end{table}

\begin{figure}[!htb]
\centering
\includegraphics[width=0.50\textwidth]{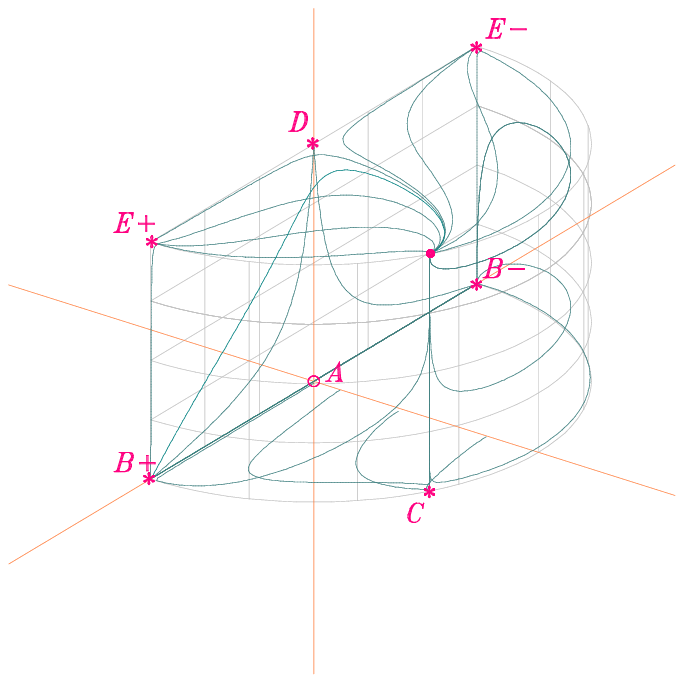}\hfill
\includegraphics[width=0.50\textwidth]{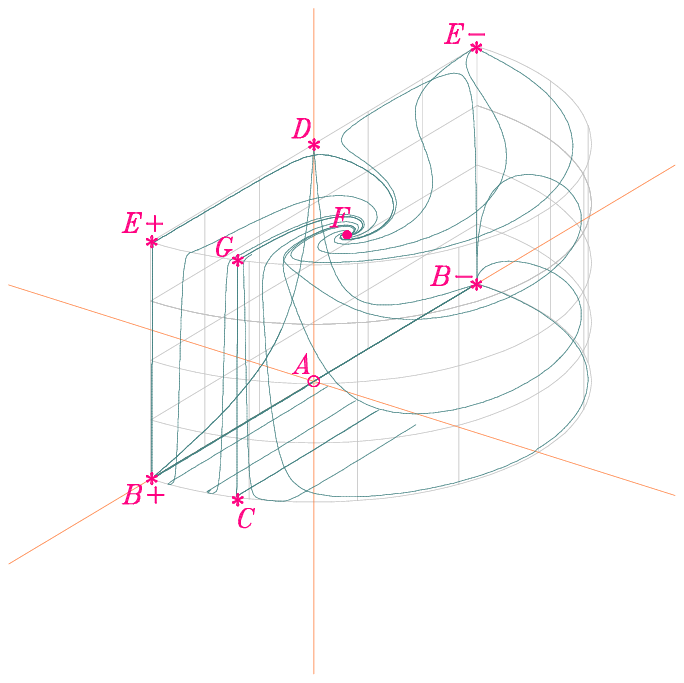}
\caption{Phase-space trajectories for model $\mathcal{C}$. The
left plot shows the stable node G, with $\lambda=1.2$ and
$\gamma=10^{-3}$. The right plot shows the stable focus F with
$\lambda=2.3$ and $\gamma=10^{-3}$. } \label{model3}
\end{figure}

This model has interesting properties. First of all, a standard matter represented by 
the unstable point $A$ exists. In particular when $\gamma>0$, the instability of the point
is generic, as it is occurs in all directions; whereas for $\gamma<0$ the point becomes  a saddle.
Secondly, an accelerated attractor exists for adequate values of the parameters. The two possible
late-time attractors as in the other two cases are found again, and the same conditions as in those
cases operate in connection with the kinematical features of the scenario they represent, if $\lambda$ is
not small enough, acceleration will not be possible and in addition dark energy will not dominate completely.

The coupling strength for this model, Eq.~(\ref{strength}) is
decreasing into the past as
\begin{equation}
  f\sim {\rho_\phi}^2/\rho_c^{1/2}\,.
\end{equation}

\subsection{Superposition of couplings}

When we combine the different coupling models together, we expect
only those critical points to be present which are critical points
of each individual model. Since we want to describe the evolution
of a universe that includes a standard matter era and evolves
towards a stable accelerating solution, we must choose the
combination of models $\mathcal{A}$ and $\mathcal{C}$, since those
are the ones that allow for a standard matter era. Therefore we
consider the composite model defined by the coupling
\begin{align}
  Q=\frac{\alpha}{H_0} \rho_\varphi^2  + \frac{\gamma}{H_0} \rho_c
  \rho_\varphi\,.
\end{align}
We note that the two couplings are decoupled in the sense that
there are no cross coupling terms in the dynamical system.

The resulting system of autonomous differential equations reads
\begin{align}
  \label{allx'}
  x'&=-3x+\lambda \frac{\sqrt{6}}{2}\,y^2+\frac{3}{2}x(1+x^2-y^2) 
  +\alpha\,\frac{3(1-z)(x^2+y^2)^2}{2xz}
  +\gamma\,\frac{3(1-z)(1-x^2-y^2)(x^2+y^2)}{2xz}\,,\\
  \label{ally'}
  y'&=-\lambda \frac{\sqrt{6}}{2}\,xy + \frac{3}{2}y(1+x^2-y^2)\,,\\
  \label{allz'}
  z'&= \frac{3}{2}z(1-z)(1+x^2-y^2)\,.
\end{align}
The critical points and their stability are listed in
Tables~\ref{models1_3} and~\ref{superstab}.

\begin{table}[!htb]
\begin{tabular}[t]{|l|c|c|c|c|c|}
\hline
Point & $x_*$ & $y_*$ & $z_*$ & Eigenvalues & $w_{tot}$\\
\hline \hline &&&&\\[-3ex]
A & $0$ & $0$ & $0$ & sgn$(\gamma) \infty \,,\frac{3}{2}\,,\frac{3}{2}$ &0\\[1ex]
\hline &&&&& \\[-3ex]
D & $0$ & $0$ & $1$ & $ -\frac{3}{2}\,,-\frac{3}{2}\,,\frac{3}{2}$ & 0\\[1ex]
\hline &&&&& \\[-3ex]
$\mathrm{E}_{\pm}$ & $\pm 1$ & $0$ & $1$ & $-3\,, 3\,,3\mp\sqrt{\frac{3}{2}}
\lambda$ & 1\\[1ex]
\hline
\hline &&&&& \\[-3ex]
F & $\frac{1}{\lambda}\sqrt{\frac{3}{2}}$ & $\frac{1}{\lambda}\sqrt{\frac{3}{2}}$
& $1$ & $-\frac{3}{2}\,, -\frac{3}{4\lambda}(\lambda\pm\sqrt{24-7\lambda^2})$ &
 0\\[1ex]
\hline &&&&& \\[-3ex]
G & $\frac{\lambda}{\sqrt{6}}$ & $\sqrt{1-\frac{\lambda^2}{6}}$ & $1$ &
$-\frac{\lambda^2}{2}\,, -3+\frac{\lambda^2}{2}\,,\lambda^2 - 3$ &
$\frac{\lambda^{2}}{2}-1$ \\[1ex]
\hline \hline
\end{tabular}
\caption{Critical points and associated eigenvalues for the
superposition of couplings for model $\mathcal{A}$ and model
$\mathcal{C}$.} \label{models1_3}
\end{table}

\begin{table}[!htb]
\begin{tabular}[t]{|l|c|c|c|c|c|c|}
\hline
Point & Stable?& $\Omega_{\phi}$ & $w_T$& Acceleration?& Existence \\
\hline \hline &&&&& \\[-3ex]
A & Unstable node for $\gamma>0$  & 0 & 0 &No& $\forall \,
\lambda,\,\alpha,\,\beta,\gamma $\\[1ex]
& Saddle node for $\gamma <0 $& & & & \\
\hline &&&&& \\[-3ex]
D & Saddle node& 0 & 0 &No& $\forall \, \lambda,\,\alpha,\,\beta,\gamma $\\[1ex]
\hline &&&&& \\[-3ex]
E$_{\pm}$ & Saddle node & 1 & 1 & No &$\forall \, \lambda,\,\alpha,\,\beta,
\, \gamma $\\[1ex]
\hline &&&&& \\[-3ex]
F & Stable focus for $\lambda^2 >\frac{24}{7}$
&$\frac{3}{\lambda^2}$ & 0&No &
$\lambda ^2 > 3$\\[1ex]
   &Stable node for $3<\lambda^2 <\frac{24}{7}$&&&& \\[1ex]
\hline &&&&&\\[-3ex]
G & Saddle node for $\lambda^2>3$ & 1& $\frac{\lambda^2}{3}-1$ &
$\lambda^2<2$ &
$\lambda^2 <6$\\[1ex]
 &Stable node for $\lambda^2<3$&&&& \\[1ex]
\hline
\end{tabular}
\caption{The properties of the critical points for the
superposition of couplings.} \label{superstab}
\end{table}

\begin{figure}[!htb]
\centering
\includegraphics[width=0.50\textwidth]{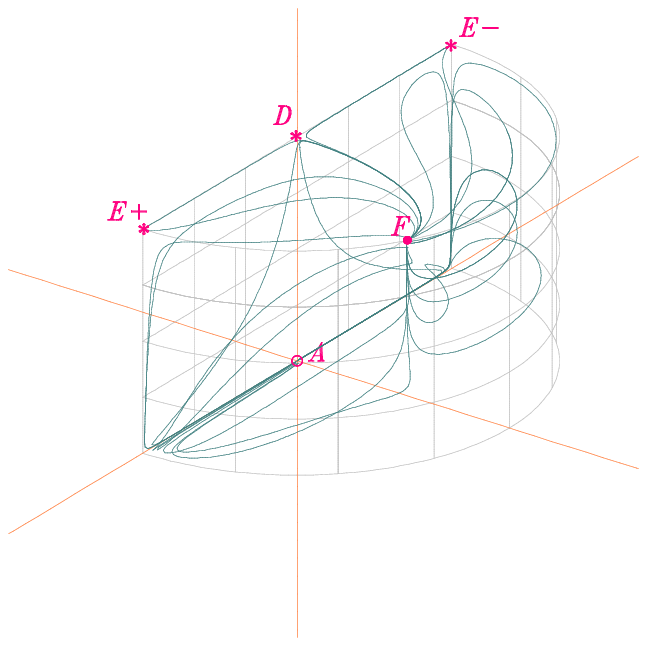}\hfill
\includegraphics[width=0.50\textwidth]{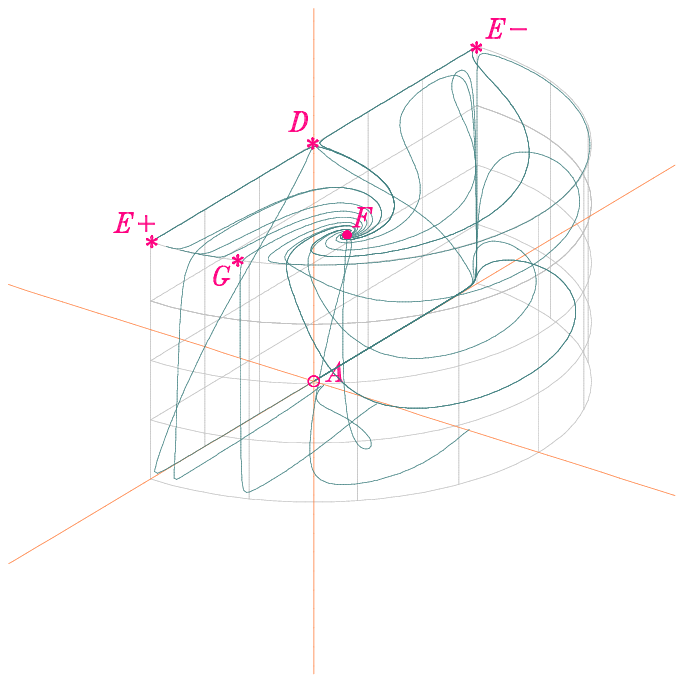}
\caption{Phase-space trajectories for the superposition of
couplings. The left plot shows the stable node G, with $\lambda=1.2$
and $\alpha= 2 \gamma=2 \times 10^{-3}$. The right plot shows the stable focus
F with $\lambda=2.3$ and $\alpha= 2 \gamma=2 \times 10^{-3}$.}
\label{super1_3}
\end{figure}

Points A and D correspond to the standard matter era and point G
is the accelerated attractor for $\lambda ^2 < 2$.

\section{Conclusions}

We have made a comprehensive analysis of the background dynamics
for a new class of models with quadratic dark sector coupling,
which are a simple physically-motivated generalization of the
linear coupling couplings with constant rates of energy transfer,
given
in~\cite{Boehmer:2008av,Valiviita:2008iv,CalderaCabral:2008bx}. The two species
which interact are dark matter and quintessence with an exponential self-interaction potential.

Of the three different terms in the general quadratic coupling, we
found that the term $Q= \mathcal B \rho_{c}^2$ leads to a universe
without a standard matter era, whereas the other two terms, $Q=
\mathcal A \rho_{\varphi}^2$ and $Q= \mathcal C
\rho_c\rho_{\varphi}^2$, do admit a standard matter era and an
evolution that connects this to a late-time
attractor. This attractor is accelerated provide the potential is flat enough. The models we have analysed provide us with the
following partial answers. These features are valid for both signs
of $\mathcal A$ or $\mathcal C$, i.e the evolution is not affected by the
direction of the energy transfer. But in the 
$\mathcal{C}>0$ case the instability of the matter era is more generic; so there is in a way more room for a transition
from the matter era to the accelerated attractor. In other words, in theses case there are less restrictions on the initial
conditions for this desired transition between asymptotic states to occur.

The critical point for late-time acceleration, G, is not a scaling
solution, since
\begin{equation}
  \Omega_{c*}=0\,, \qquad \Omega_{\varphi *}=1\,.
\end{equation}
This is similar to the asymptotic behaviour of the standard
$\Lambda$CDM model. Thus the quadratic models do not produce a
constant non-zero and finite ratio $\Omega_{c*}/\Omega_{\phi *}$,
and therefore do not address the coincidence problem in this
sense. The linear coupling $Q=\Gamma \rho_c$ leads the same
critical point G; this accelerated solution is an attractor when
$\Gamma>0$, i.e when dark matter is decaying into dark energy.

The quadratic models which admit a viable background evolution can
be compared to observations in order to constrain the parameters
$\alpha$ and $\gamma$. This will require an investigation of the
cosmological perturbations in these models.

\acknowledgements

We thank Jussi V\"aliviita for useful discussions. The work of RM is supported by the UK's Science \& Technology Facilities Council. GCC is supported by the Programme Alban (the European Union Programme of High Level Scholarships for Latin America), scholarship No. E06D103604MX, and the Mexican National Council for Science and Technology, CONACYT, scholarship No. 192680.                                            The work of R.L. is supported by the University of the Basque Country through
research grant GIU06/37 and by the Spanish Ministry of
Science and Innovation through research grant FIS2007-
61800.

\bibliographystyle{apsrev}
\bibliography{biblio_quad}

\begin{thebibliography}{49}
\expandafter\ifx\csname natexlab\endcsname\relax\def\natexlab#1{#1}\fi
\expandafter\ifx\csname bibnamefont\endcsname\relax
  \def\bibnamefont#1{#1}\fi
\expandafter\ifx\csname bibfnamefont\endcsname\relax
  \def\bibfnamefont#1{#1}\fi
\expandafter\ifx\csname citenamefont\endcsname\relax
  \def\citenamefont#1{#1}\fi
\expandafter\ifx\csname url\endcsname\relax
  \def\url#1{\texttt{#1}}\fi
\expandafter\ifx\csname urlprefix\endcsname\relax\def\urlprefix{URL }\fi
\providecommand{\bibinfo}[2]{#2}
\providecommand{\eprint}[2][]{\url{#2}}

\bibitem[{\citenamefont{Dunkley et~al.}(2008)}]{Dunkley:2008ie}
\bibinfo{author}{\bibfnamefont{J.}~\bibnamefont{Dunkley}} \bibnamefont{et~al.}
  (\bibinfo{collaboration}{WMAP}) (\bibinfo{year}{2008}), \eprint{0803.0586}.

\bibitem[{\citenamefont{Collaboration}(2009)}]{Collaboration:2009nf}
\bibinfo{author}{\bibfnamefont{T.~I.} \bibnamefont{Collaboration}}
  (\bibinfo{year}{2009}), \eprint{0902.0675}.

\bibitem[{\citenamefont{Amendola and
  Tocchini-Valentini}(2002)}]{Amendola:2001rc}
\bibinfo{author}{\bibfnamefont{L.}~\bibnamefont{Amendola}} \bibnamefont{and}
  \bibinfo{author}{\bibfnamefont{D.}~\bibnamefont{Tocchini-Valentini}},
  \bibinfo{journal}{Phys. Rev.} \textbf{\bibinfo{volume}{D66}},
  \bibinfo{pages}{043528} (\bibinfo{year}{2002}), \eprint{astro-ph/0111535}.

\bibitem[{\citenamefont{Koyama et~al.}(2009)\citenamefont{Koyama, Maartens, and
  Song}}]{Koyama:2009gd}
\bibinfo{author}{\bibfnamefont{K.}~\bibnamefont{Koyama}},
  \bibinfo{author}{\bibfnamefont{R.}~\bibnamefont{Maartens}}, \bibnamefont{and}
  \bibinfo{author}{\bibfnamefont{Y.-S.} \bibnamefont{Song}}
  (\bibinfo{year}{2009}), \eprint{0907.2126}.

\bibitem[{\citenamefont{Wetterich}(1995)}]{Wetterich:1994bg}
\bibinfo{author}{\bibfnamefont{C.}~\bibnamefont{Wetterich}},
  \bibinfo{journal}{Astron. Astrophys.} \textbf{\bibinfo{volume}{301}},
  \bibinfo{pages}{321} (\bibinfo{year}{1995}), \eprint{hep-th/9408025}.

\bibitem[{\citenamefont{Amendola}(1999)}]{Amendola:1999qq}
\bibinfo{author}{\bibfnamefont{L.}~\bibnamefont{Amendola}},
  \bibinfo{journal}{Phys. Rev.} \textbf{\bibinfo{volume}{D60}},
  \bibinfo{pages}{043501} (\bibinfo{year}{1999}), \eprint{astro-ph/9904120}.

\bibitem[{\citenamefont{Billyard and Coley}(2000)}]{Billyard:2000bh}
\bibinfo{author}{\bibfnamefont{A.~P.} \bibnamefont{Billyard}} \bibnamefont{and}
  \bibinfo{author}{\bibfnamefont{A.~A.} \bibnamefont{Coley}},
  \bibinfo{journal}{Phys. Rev.} \textbf{\bibinfo{volume}{D61}},
  \bibinfo{pages}{083503} (\bibinfo{year}{2000}), \eprint{astro-ph/9908224}.

\bibitem[{\citenamefont{Zimdahl and Pavon}(2001)}]{Zimdahl:2001ar}
\bibinfo{author}{\bibfnamefont{W.}~\bibnamefont{Zimdahl}} \bibnamefont{and}
  \bibinfo{author}{\bibfnamefont{D.}~\bibnamefont{Pavon}},
  \bibinfo{journal}{Phys. Lett.} \textbf{\bibinfo{volume}{B521}},
  \bibinfo{pages}{133} (\bibinfo{year}{2001}), \eprint{astro-ph/0105479}.

\bibitem[{\citenamefont{Farrar and Peebles}(2004)}]{Farrar:2003uw}
\bibinfo{author}{\bibfnamefont{G.~R.} \bibnamefont{Farrar}} \bibnamefont{and}
  \bibinfo{author}{\bibfnamefont{P.~J.~E.} \bibnamefont{Peebles}},
  \bibinfo{journal}{Astrophys. J.} \textbf{\bibinfo{volume}{604}},
  \bibinfo{pages}{1} (\bibinfo{year}{2004}), \eprint{astro-ph/0307316}.

\bibitem[{\citenamefont{Chimento et~al.}(2003)\citenamefont{Chimento, Jakubi,
  Pavon, and Zimdahl}}]{Chimento:2003iea}
\bibinfo{author}{\bibfnamefont{L.~P.} \bibnamefont{Chimento}},
  \bibinfo{author}{\bibfnamefont{A.~S.} \bibnamefont{Jakubi}},
  \bibinfo{author}{\bibfnamefont{D.}~\bibnamefont{Pavon}}, \bibnamefont{and}
  \bibinfo{author}{\bibfnamefont{W.}~\bibnamefont{Zimdahl}},
  \bibinfo{journal}{Phys. Rev.} \textbf{\bibinfo{volume}{D67}},
  \bibinfo{pages}{083513} (\bibinfo{year}{2003}), \eprint{astro-ph/0303145}.

\bibitem[{\citenamefont{Olivares et~al.}(2005)\citenamefont{Olivares,
  Atrio-Barandela, and Pavon}}]{Olivares:2005tb}
\bibinfo{author}{\bibfnamefont{G.}~\bibnamefont{Olivares}},
  \bibinfo{author}{\bibfnamefont{F.}~\bibnamefont{Atrio-Barandela}},
  \bibnamefont{and} \bibinfo{author}{\bibfnamefont{D.}~\bibnamefont{Pavon}},
  \bibinfo{journal}{Phys. Rev.} \textbf{\bibinfo{volume}{D71}},
  \bibinfo{pages}{063523} (\bibinfo{year}{2005}), \eprint{astro-ph/0503242}.

\bibitem[{\citenamefont{Sadjadi and Alimohammadi}(2006)}]{Sadjadi:2006qp}
\bibinfo{author}{\bibfnamefont{H.~M.} \bibnamefont{Sadjadi}} \bibnamefont{and}
  \bibinfo{author}{\bibfnamefont{M.}~\bibnamefont{Alimohammadi}},
  \bibinfo{journal}{Phys. Rev.} \textbf{\bibinfo{volume}{D74}},
  \bibinfo{pages}{103007} (\bibinfo{year}{2006}), \eprint{gr-qc/0610080}.

\bibitem[{\citenamefont{Guo et~al.}(2007)\citenamefont{Guo, Ohta, and
  Tsujikawa}}]{Guo:2007zk}
\bibinfo{author}{\bibfnamefont{Z.-K.} \bibnamefont{Guo}},
  \bibinfo{author}{\bibfnamefont{N.}~\bibnamefont{Ohta}}, \bibnamefont{and}
  \bibinfo{author}{\bibfnamefont{S.}~\bibnamefont{Tsujikawa}},
  \bibinfo{journal}{Phys. Rev.} \textbf{\bibinfo{volume}{D76}},
  \bibinfo{pages}{023508} (\bibinfo{year}{2007}), \eprint{astro-ph/0702015}.

\bibitem[{\citenamefont{Kim et~al.}(2007)\citenamefont{Kim, Lee, and
  Myung}}]{Kim:2007dp}
\bibinfo{author}{\bibfnamefont{K.~Y.} \bibnamefont{Kim}},
  \bibinfo{author}{\bibfnamefont{H.~W.} \bibnamefont{Lee}}, \bibnamefont{and}
  \bibinfo{author}{\bibfnamefont{Y.~S.} \bibnamefont{Myung}},
  \bibinfo{journal}{Mod. Phys. Lett.} \textbf{\bibinfo{volume}{A22}},
  \bibinfo{pages}{2631} (\bibinfo{year}{2007}), \eprint{0706.2444}.

\bibitem[{\citenamefont{Boehmer et~al.}(2008)\citenamefont{Boehmer,
  Caldera-Cabral, Lazkoz, and Maartens}}]{Boehmer:2008av}
\bibinfo{author}{\bibfnamefont{C.~G.} \bibnamefont{Boehmer}},
  \bibinfo{author}{\bibfnamefont{G.}~\bibnamefont{Caldera-Cabral}},
  \bibinfo{author}{\bibfnamefont{R.}~\bibnamefont{Lazkoz}}, \bibnamefont{and}
  \bibinfo{author}{\bibfnamefont{R.}~\bibnamefont{Maartens}},
  \bibinfo{journal}{Phys. Rev.} \textbf{\bibinfo{volume}{D78}},
  \bibinfo{pages}{023505} (\bibinfo{year}{2008}), \eprint{0801.1565}.

\bibitem[{\citenamefont{He and Wang}(2008)}]{He:2008tn}
\bibinfo{author}{\bibfnamefont{J.-H.} \bibnamefont{He}} \bibnamefont{and}
  \bibinfo{author}{\bibfnamefont{B.}~\bibnamefont{Wang}},
  \bibinfo{journal}{JCAP} \textbf{\bibinfo{volume}{0806}}, \bibinfo{pages}{010}
  (\bibinfo{year}{2008}), \eprint{0801.4233}.

\bibitem[{\citenamefont{Chen et~al.}(2008)\citenamefont{Chen, Wang, and
  Jing}}]{Chen:2008ca}
\bibinfo{author}{\bibfnamefont{S.}~\bibnamefont{Chen}},
  \bibinfo{author}{\bibfnamefont{B.}~\bibnamefont{Wang}}, \bibnamefont{and}
  \bibinfo{author}{\bibfnamefont{J.}~\bibnamefont{Jing}},
  \bibinfo{journal}{Phys. Rev.} \textbf{\bibinfo{volume}{D78}},
  \bibinfo{pages}{123503} (\bibinfo{year}{2008}), \eprint{0808.3482}.

\bibitem[{\citenamefont{Quartin et~al.}(2008)\citenamefont{Quartin, Calvao,
  Joras, Reis, and Waga}}]{Quartin:2008px}
\bibinfo{author}{\bibfnamefont{M.}~\bibnamefont{Quartin}},
  \bibinfo{author}{\bibfnamefont{M.~O.} \bibnamefont{Calvao}},
  \bibinfo{author}{\bibfnamefont{S.~E.} \bibnamefont{Joras}},
  \bibinfo{author}{\bibfnamefont{R.~R.~R.} \bibnamefont{Reis}},
  \bibnamefont{and} \bibinfo{author}{\bibfnamefont{I.}~\bibnamefont{Waga}},
  \bibinfo{journal}{JCAP} \textbf{\bibinfo{volume}{0805}}, \bibinfo{pages}{007}
  (\bibinfo{year}{2008}), \eprint{0802.0546}.

\bibitem[{\citenamefont{Pereira and Jesus}(2009)}]{Pereira:2008at}
\bibinfo{author}{\bibfnamefont{S.~H.} \bibnamefont{Pereira}} \bibnamefont{and}
  \bibinfo{author}{\bibfnamefont{J.~F.} \bibnamefont{Jesus}},
  \bibinfo{journal}{Phys. Rev.} \textbf{\bibinfo{volume}{D79}},
  \bibinfo{pages}{043517} (\bibinfo{year}{2009}), \eprint{0811.0099}.

\bibitem[{\citenamefont{Quercellini et~al.}(2008)\citenamefont{Quercellini,
  Bruni, Balbi, and Pietrobon}}]{Quercellini:2008vh}
\bibinfo{author}{\bibfnamefont{C.}~\bibnamefont{Quercellini}},
  \bibinfo{author}{\bibfnamefont{M.}~\bibnamefont{Bruni}},
  \bibinfo{author}{\bibfnamefont{A.}~\bibnamefont{Balbi}}, \bibnamefont{and}
  \bibinfo{author}{\bibfnamefont{D.}~\bibnamefont{Pietrobon}}
  (\bibinfo{year}{2008}), \eprint{0803.1976}.

\bibitem[{\citenamefont{Valiviita et~al.}(2008)\citenamefont{Valiviita,
  Majerotto, and Maartens}}]{Valiviita:2008iv}
\bibinfo{author}{\bibfnamefont{J.}~\bibnamefont{Valiviita}},
  \bibinfo{author}{\bibfnamefont{E.}~\bibnamefont{Majerotto}},
  \bibnamefont{and} \bibinfo{author}{\bibfnamefont{R.}~\bibnamefont{Maartens}},
  \bibinfo{journal}{JCAP} \textbf{\bibinfo{volume}{0807}}, \bibinfo{pages}{020}
  (\bibinfo{year}{2008}), \eprint{0804.0232}.

\bibitem[{\citenamefont{Amendola et~al.}(2003)\citenamefont{Amendola,
  Quercellini, Tocchini-Valentini, and Pasqui}}]{Amendola:2002bs}
\bibinfo{author}{\bibfnamefont{L.}~\bibnamefont{Amendola}},
  \bibinfo{author}{\bibfnamefont{C.}~\bibnamefont{Quercellini}},
  \bibinfo{author}{\bibfnamefont{D.}~\bibnamefont{Tocchini-Valentini}},
  \bibnamefont{and} \bibinfo{author}{\bibfnamefont{A.}~\bibnamefont{Pasqui}},
  \bibinfo{journal}{Astrophys. J.} \textbf{\bibinfo{volume}{583}},
  \bibinfo{pages}{L53} (\bibinfo{year}{2003}), \eprint{astro-ph/0205097}.

\bibitem[{\citenamefont{Koivisto}(2005)}]{Koivisto:2005nr}
\bibinfo{author}{\bibfnamefont{T.}~\bibnamefont{Koivisto}},
  \bibinfo{journal}{Phys. Rev.} \textbf{\bibinfo{volume}{D72}},
  \bibinfo{pages}{043516} (\bibinfo{year}{2005}), \eprint{astro-ph/0504571}.

\bibitem[{\citenamefont{Olivares et~al.}(2006)\citenamefont{Olivares,
  Atrio-Barandela, and Pavon}}]{Olivares:2006jr}
\bibinfo{author}{\bibfnamefont{G.}~\bibnamefont{Olivares}},
  \bibinfo{author}{\bibfnamefont{F.}~\bibnamefont{Atrio-Barandela}},
  \bibnamefont{and} \bibinfo{author}{\bibfnamefont{D.}~\bibnamefont{Pavon}},
  \bibinfo{journal}{Phys. Rev.} \textbf{\bibinfo{volume}{D74}},
  \bibinfo{pages}{043521} (\bibinfo{year}{2006}), \eprint{astro-ph/0607604}.

\bibitem[{\citenamefont{Mainini and Bonometto}(2007)}]{Mainini:2007ft}
\bibinfo{author}{\bibfnamefont{R.}~\bibnamefont{Mainini}} \bibnamefont{and}
  \bibinfo{author}{\bibfnamefont{S.}~\bibnamefont{Bonometto}},
  \bibinfo{journal}{JCAP} \textbf{\bibinfo{volume}{0706}}, \bibinfo{pages}{020}
  (\bibinfo{year}{2007}), \eprint{astro-ph/0703303}.

\bibitem[{\citenamefont{Bean et~al.}(2008{\natexlab{a}})\citenamefont{Bean,
  Flanagan, and Trodden}}]{Bean:2007ny}
\bibinfo{author}{\bibfnamefont{R.}~\bibnamefont{Bean}},
  \bibinfo{author}{\bibfnamefont{E.~E.} \bibnamefont{Flanagan}},
  \bibnamefont{and} \bibinfo{author}{\bibfnamefont{M.}~\bibnamefont{Trodden}},
  \bibinfo{journal}{New J. Phys.} \textbf{\bibinfo{volume}{10}},
  \bibinfo{pages}{033006} (\bibinfo{year}{2008}{\natexlab{a}}),
  \eprint{0709.1124}.

\bibitem[{\citenamefont{Vergani et~al.}(2008)\citenamefont{Vergani, Colombo,
  La~Vacca, and Bonometto}}]{Vergani:2008jv}
\bibinfo{author}{\bibfnamefont{L.}~\bibnamefont{Vergani}},
  \bibinfo{author}{\bibfnamefont{L.~P.~L.} \bibnamefont{Colombo}},
  \bibinfo{author}{\bibfnamefont{G.}~\bibnamefont{La~Vacca}}, \bibnamefont{and}
  \bibinfo{author}{\bibfnamefont{S.~A.} \bibnamefont{Bonometto}}
  (\bibinfo{year}{2008}), \eprint{0804.0285}.

\bibitem[{\citenamefont{Pettorino and Baccigalupi}(2008)}]{Pettorino:2008ez}
\bibinfo{author}{\bibfnamefont{V.}~\bibnamefont{Pettorino}} \bibnamefont{and}
  \bibinfo{author}{\bibfnamefont{C.}~\bibnamefont{Baccigalupi}},
  \bibinfo{journal}{Phys. Rev.} \textbf{\bibinfo{volume}{D77}},
  \bibinfo{pages}{103003} (\bibinfo{year}{2008}), \eprint{0802.1086}.

\bibitem[{\citenamefont{Schaefer}(2008)}]{Schaefer:2008qs}
\bibinfo{author}{\bibfnamefont{B.~M.} \bibnamefont{Schaefer}}
  (\bibinfo{year}{2008}), \eprint{0803.2239}.

\bibitem[{\citenamefont{Schaefer et~al.}(2008)\citenamefont{Schaefer,
  Caldera-Cabral, and Maartens}}]{Schaefer:2008ku}
\bibinfo{author}{\bibfnamefont{B.~M.} \bibnamefont{Schaefer}},
  \bibinfo{author}{\bibfnamefont{G.~A.} \bibnamefont{Caldera-Cabral}},
  \bibnamefont{and} \bibinfo{author}{\bibfnamefont{R.}~\bibnamefont{Maartens}}
  (\bibinfo{year}{2008}), \eprint{0803.2154}.

\bibitem[{\citenamefont{La~Vacca and Colombo}(2008)}]{LaVacca:2008kq}
\bibinfo{author}{\bibfnamefont{G.}~\bibnamefont{La~Vacca}} \bibnamefont{and}
  \bibinfo{author}{\bibfnamefont{L.~P.~L.} \bibnamefont{Colombo}},
  \bibinfo{journal}{JCAP} \textbf{\bibinfo{volume}{0804}}, \bibinfo{pages}{007}
  (\bibinfo{year}{2008}), \eprint{0803.1640}.

\bibitem[{\citenamefont{He et~al.}(2009{\natexlab{a}})\citenamefont{He, Wang,
  and Abdalla}}]{He:2008si}
\bibinfo{author}{\bibfnamefont{J.-H.} \bibnamefont{He}},
  \bibinfo{author}{\bibfnamefont{B.}~\bibnamefont{Wang}}, \bibnamefont{and}
  \bibinfo{author}{\bibfnamefont{E.}~\bibnamefont{Abdalla}},
  \bibinfo{journal}{Phys. Lett.} \textbf{\bibinfo{volume}{B671}},
  \bibinfo{pages}{139} (\bibinfo{year}{2009}{\natexlab{a}}),
  \eprint{0807.3471}.

\bibitem[{\citenamefont{Bean et~al.}(2008{\natexlab{b}})\citenamefont{Bean,
  Flanagan, Laszlo, and Trodden}}]{Bean:2008ac}
\bibinfo{author}{\bibfnamefont{R.}~\bibnamefont{Bean}},
  \bibinfo{author}{\bibfnamefont{E.~E.} \bibnamefont{Flanagan}},
  \bibinfo{author}{\bibfnamefont{I.}~\bibnamefont{Laszlo}}, \bibnamefont{and}
  \bibinfo{author}{\bibfnamefont{M.}~\bibnamefont{Trodden}},
  \bibinfo{journal}{Phys. Rev.} \textbf{\bibinfo{volume}{D78}},
  \bibinfo{pages}{123514} (\bibinfo{year}{2008}{\natexlab{b}}),
  \eprint{0808.1105}.

\bibitem[{\citenamefont{Corasaniti}(2008)}]{Corasaniti:2008kx}
\bibinfo{author}{\bibfnamefont{P.~S.} \bibnamefont{Corasaniti}},
  \bibinfo{journal}{Phys. Rev.} \textbf{\bibinfo{volume}{D78}},
  \bibinfo{pages}{083538} (\bibinfo{year}{2008}), \eprint{0808.1646}.

\bibitem[{\citenamefont{Chongchitnan}(2009)}]{Chongchitnan:2008ry}
\bibinfo{author}{\bibfnamefont{S.}~\bibnamefont{Chongchitnan}},
  \bibinfo{journal}{Phys. Rev.} \textbf{\bibinfo{volume}{D79}},
  \bibinfo{pages}{043522} (\bibinfo{year}{2009}), \eprint{0810.5411}.

\bibitem[{\citenamefont{Jackson et~al.}(2009)\citenamefont{Jackson, Taylor, and
  Berera}}]{Jackson:2009mz}
\bibinfo{author}{\bibfnamefont{B.~M.} \bibnamefont{Jackson}},
  \bibinfo{author}{\bibfnamefont{A.}~\bibnamefont{Taylor}}, \bibnamefont{and}
  \bibinfo{author}{\bibfnamefont{A.}~\bibnamefont{Berera}},
  \bibinfo{journal}{Phys. Rev.} \textbf{\bibinfo{volume}{D79}},
  \bibinfo{pages}{043526} (\bibinfo{year}{2009}), \eprint{0901.3272}.

\bibitem[{\citenamefont{Gavela et~al.}(2009)\citenamefont{Gavela, Hernandez,
  Honorez, Mena, and Rigolin}}]{Gavela:2009cy}
\bibinfo{author}{\bibfnamefont{M.~B.} \bibnamefont{Gavela}},
  \bibinfo{author}{\bibfnamefont{D.}~\bibnamefont{Hernandez}},
  \bibinfo{author}{\bibfnamefont{L.~L.} \bibnamefont{Honorez}},
  \bibinfo{author}{\bibfnamefont{O.}~\bibnamefont{Mena}}, \bibnamefont{and}
  \bibinfo{author}{\bibfnamefont{S.}~\bibnamefont{Rigolin}}
  (\bibinfo{year}{2009}), \eprint{0901.1611}.

\bibitem[{\citenamefont{La~Vacca et~al.}(2009)\citenamefont{La~Vacca,
  Kristiansen, Colombo, Mainini, and Bonometto}}]{LaVacca:2009yp}
\bibinfo{author}{\bibfnamefont{G.}~\bibnamefont{La~Vacca}},
  \bibinfo{author}{\bibfnamefont{J.~R.} \bibnamefont{Kristiansen}},
  \bibinfo{author}{\bibfnamefont{L.~P.~L.} \bibnamefont{Colombo}},
  \bibinfo{author}{\bibfnamefont{R.}~\bibnamefont{Mainini}}, \bibnamefont{and}
  \bibinfo{author}{\bibfnamefont{S.~A.} \bibnamefont{Bonometto}},
  \bibinfo{journal}{JCAP} \textbf{\bibinfo{volume}{0904}}, \bibinfo{pages}{007}
  (\bibinfo{year}{2009}), \eprint{0902.2711}.

\bibitem[{\citenamefont{He et~al.}(2009{\natexlab{b}})\citenamefont{He, Wang,
  and Jing}}]{He:2009mz}
\bibinfo{author}{\bibfnamefont{J.-H.} \bibnamefont{He}},
  \bibinfo{author}{\bibfnamefont{B.}~\bibnamefont{Wang}}, \bibnamefont{and}
  \bibinfo{author}{\bibfnamefont{Y.~P.} \bibnamefont{Jing}},
  \bibinfo{journal}{JCAP} \textbf{\bibinfo{volume}{0907}}, \bibinfo{pages}{030}
  (\bibinfo{year}{2009}{\natexlab{b}}), \eprint{0902.0660}.

\bibitem[{\citenamefont{Caldera-Cabral
  et~al.}(2009{\natexlab{a}})\citenamefont{Caldera-Cabral, Maartens, and
  Schaefer}}]{CalderaCabral:2009ja}
\bibinfo{author}{\bibfnamefont{G.}~\bibnamefont{Caldera-Cabral}},
  \bibinfo{author}{\bibfnamefont{R.}~\bibnamefont{Maartens}}, \bibnamefont{and}
  \bibinfo{author}{\bibfnamefont{B.~M.} \bibnamefont{Schaefer}},
  \bibinfo{journal}{JCAP} \textbf{\bibinfo{volume}{0907}}, \bibinfo{pages}{027}
  (\bibinfo{year}{2009}{\natexlab{a}}), \eprint{0905.0492}.

\bibitem[{\citenamefont{He et~al.}(2009{\natexlab{c}})\citenamefont{He, Wang,
  and Zhang}}]{He:2009pd}
\bibinfo{author}{\bibfnamefont{J.-H.} \bibnamefont{He}},
  \bibinfo{author}{\bibfnamefont{B.}~\bibnamefont{Wang}}, \bibnamefont{and}
  \bibinfo{author}{\bibfnamefont{P.}~\bibnamefont{Zhang}}
  (\bibinfo{year}{2009}{\natexlab{c}}), \eprint{0906.0677}.

\bibitem[{\citenamefont{Valiviita et~al.}(2009)\citenamefont{Valiviita,
  Maartens, and Majerotto}}]{Valiviita:2009nu}
\bibinfo{author}{\bibfnamefont{J.}~\bibnamefont{Valiviita}},
  \bibinfo{author}{\bibfnamefont{R.}~\bibnamefont{Maartens}}, \bibnamefont{and}
  \bibinfo{author}{\bibfnamefont{E.}~\bibnamefont{Majerotto}}
  (\bibinfo{year}{2009}), \eprint{0907.4987}.

\bibitem[{\citenamefont{Majerotto et~al.}(2009)\citenamefont{Majerotto,
  Valiviita, and Maartens}}]{Majerotto:2009np}
\bibinfo{author}{\bibfnamefont{E.}~\bibnamefont{Majerotto}},
  \bibinfo{author}{\bibfnamefont{J.}~\bibnamefont{Valiviita}},
  \bibnamefont{and} \bibinfo{author}{\bibfnamefont{R.}~\bibnamefont{Maartens}}
  (\bibinfo{year}{2009}), \eprint{0907.4981}.

\bibitem[{\citenamefont{Turner}(1983)}]{Turner:1983he}
\bibinfo{author}{\bibfnamefont{M.~S.} \bibnamefont{Turner}},
  \bibinfo{journal}{Phys. Rev.} \textbf{\bibinfo{volume}{D28}},
  \bibinfo{pages}{1243} (\bibinfo{year}{1983}).

\bibitem[{\citenamefont{Cen}(2000)}]{Cen:2000xv}
\bibinfo{author}{\bibfnamefont{R.}~\bibnamefont{Cen}} (\bibinfo{year}{2000}),
  \eprint{astro-ph/0005206}.

\bibitem[{\citenamefont{Malik et~al.}(2003)\citenamefont{Malik, Wands, and
  Ungarelli}}]{Malik:2002jb}
\bibinfo{author}{\bibfnamefont{K.~A.} \bibnamefont{Malik}},
  \bibinfo{author}{\bibfnamefont{D.}~\bibnamefont{Wands}}, \bibnamefont{and}
  \bibinfo{author}{\bibfnamefont{C.}~\bibnamefont{Ungarelli}},
  \bibinfo{journal}{Phys. Rev.} \textbf{\bibinfo{volume}{D67}},
  \bibinfo{pages}{063516} (\bibinfo{year}{2003}), \eprint{astro-ph/0211602}.

\bibitem[{\citenamefont{Ziaeepour}(2004)}]{Ziaeepour:2003qs}
\bibinfo{author}{\bibfnamefont{H.}~\bibnamefont{Ziaeepour}},
  \bibinfo{journal}{Phys. Rev.} \textbf{\bibinfo{volume}{D69}},
  \bibinfo{pages}{063512} (\bibinfo{year}{2004}), \eprint{astro-ph/0308515}.

\bibitem[{\citenamefont{Copeland et~al.}(1998)\citenamefont{Copeland, Liddle,
  and Wands}}]{Copeland:1997et}
\bibinfo{author}{\bibfnamefont{E.~J.} \bibnamefont{Copeland}},
  \bibinfo{author}{\bibfnamefont{A.~R.} \bibnamefont{Liddle}},
  \bibnamefont{and} \bibinfo{author}{\bibfnamefont{D.}~\bibnamefont{Wands}},
  \bibinfo{journal}{Phys. Rev.} \textbf{\bibinfo{volume}{D57}},
  \bibinfo{pages}{4686} (\bibinfo{year}{1998}), \eprint{gr-qc/9711068}.

\bibitem[{\citenamefont{Caldera-Cabral
  et~al.}(2009{\natexlab{b}})\citenamefont{Caldera-Cabral, Maartens, and
  Urena-Lopez}}]{CalderaCabral:2008bx}
\bibinfo{author}{\bibfnamefont{G.}~\bibnamefont{Caldera-Cabral}},
  \bibinfo{author}{\bibfnamefont{R.}~\bibnamefont{Maartens}}, \bibnamefont{and}
  \bibinfo{author}{\bibfnamefont{L.~A.} \bibnamefont{Urena-Lopez}},
  \bibinfo{journal}{Phys. Rev.} \textbf{\bibinfo{volume}{D79}},
  \bibinfo{pages}{063518} (\bibinfo{year}{2009}{\natexlab{b}}),
  \eprint{0812.1827}.

\end{thebibliography}

\end{document}